%% file: paper_revised.tex
\documentclass[11pt]{article}
\usepackage{amsmath,amssymb}
\usepackage[dvipdfm]{graphicx}
\usepackage{color}

\newcommand{\ba}{\begin{alignat}{3}}

\begin{document}

\begin{flushright}
OU-HET 779
\\
\today
\end{flushright}
\vskip1cm
\begin{center}
{\LARGE {\bf 
Position space formulation for Dirac fermions on  honeycomb lattice}}
\vskip3cm
{\large 
{\bf Masaki Hirotsu$^{1}$ \footnote{hirotsu@hetmail.phys.sci.osaka-u.ac.jp}, 
Tetsuya Onogi$^{1}$ \footnote{onogi@phys.sci.osaka-u.ac.jp}, 
and Eigo Shintani$^{2,3}$ \footnote{shintani@kph.uni-mainz.jp}
}

\vskip1cm
\it $^{1}$Department of Physics, Graduate School of Science, 
\\
Osaka University, Toyonaka, Osaka 560-0043, Japan
\\
$^{2}$PRISMA Cluster of Excellence,
Institut f{\"u}r Kernphysik and Helmholtz Institute Mainz,
Johannes Gutenberg-Universit{\"a}t Mainz, D-55099 Mainz, Germany
$^{3}$RIKEN-BNL Research Center, Brookhaven National Laboratory, Upton, NY 11973, USA
}

\end{center}

\vskip1cm
\begin{abstract}
We study how to construct Dirac fermion defined on the honeycomb lattice in position space.
Starting from the nearest neighbor interaction in tight binding model, 
we show that the Hamiltonian is constructed by 
kinetic term and second derivative term of three flavor Dirac fermions 
in which one flavor has a mass of cutoff order and the other flavors are massless. 
In this formulation, the structure of the Dirac point is simplified
so that its uniqueness can be easily shown even if we consider the 
next-to-nearest neighbor interaction.
We also show that there is a hidden exact $U(1)$ symmetry (flavor-chiral symmetry)
at finite lattice spacing, which protects the masslessness of the Dirac fermion,
and discuss the analogy with the staggered fermion formulation.
\end{abstract}

\vfill\eject

\section{Introduction}
%
Graphene forms from a layer of carbon atoms with hexagonal tiling 
\cite{Novoselov et al.,Novoselov:2005kj,Zhang:2005zz,M. I. Katsnelson} 
and it is much discussed in condensed matter physics
as well as high energy physics for its remarkable features 
(see \cite{CastroNeto:2009zz,Kotov:2010yh} and references therein).
One of the most important features of Graphene is 
that the quasiparticles behave like massless Dirac fermion
with effective speed of light near $c/300$ \cite{Novoselov:2005kj,D. C. Elias}.
An explanation to the question why massless Dirac fermion emerges 
in non-relativistic many body system
was primarily given by Semenoff \cite{Semenoff:1984dq}.
In this model, the low energy excitations  
around two independent Dirac points on the fermi-surface is described 
by two relativistic Weyl fermions having opposite chiralities, 
which are also regarded as massless Dirac fermion.

Although Semenoff's model is remarkable for its peculiar feature, it is not the first case with exact massless Dirac fermion from the lattice. 
In lattice gauge theory, there are several formulations to describe Dirac fermion on the lattice.
In the staggered fermion formulation \cite{Susskind:1976jm},
the $2^{d-2}$flavor Dirac fermions emerge at low energy from a single spinless 
fermion hopping around the d-dimensional hypercubic lattice, in close analogy to the Semenoff's model. 
The  emergence of the Dirac fermion in staggered fermion has been studied
in momentum space \cite{Sharatchandra:1981si} and 
in position space \cite{KlubergStern:1983dg}. 
In the former case, 
the fermion field is divided into  $2^d$ components 
corresponding to the subdomains in the total momentum space. 
In the latter case,  $2^{d}$ spin-flavor degrees of freedom of the Dirac fermion 
arise from the sites within the d-dimensional hypercubic  unit cell. 

In the case of honeycomb lattice in 2+1 dimension,
Dirac fermion field has been defined as two excitations 
on different regions of Brillouin Zone (BZ) in the continuum space-time
\cite{Semenoff:1984dq}. Since this approach is very similar to the 
momentum space formulation  for staggered fermion, it is natural to expect 
that  position space formulation might also be possible for Graphene model. 
Since the position space formulation easily extends local gauge interacting theory, 
it enables us to implement the dynamical calculation of physical observables
in Monte-Carlo simulation more straightforwardly
\cite{Brower:2012zd,Buividovich:2012nx}.
Furthermore, this formulation also has the manifest structure of flavor symmetry 
of Dirac fermion field, and so that 
the quantum number of low energy excitations is clearly identified.

In this paper, we show how to construct the Dirac fermion in
position space on honeycomb lattice.
It may be useful to advance a study of the dynamical nature of Graphene 
with numerical approaches using Monte-Carlo simulation 
\cite{Drut:2009,Armour:2010,Buividovich:2012uk,Shintani:2012sz}.
This approach plays an important role for more rigorous discussion
for than modeling one \cite{Herbut:2006cs,Son:2007} (also see \cite{Semenoff:2011jf}).

This paper is organized as follows.
In section \ref{sec:conv}, we briefly review a momentum space formulation 
of Dirac fermion derived from tight-binding approximation of 
Graphene model.
In section \ref{sec:form}, after introducing the formulation in position space, 
we discuss uniqueness of Dirac point and existence of physical mode, and then
in section \ref{sec:continuum},
we show that two massless Dirac fermions appear at low energy region. 
In section \ref{sec:chiral}, we discuss the exact flavor-chiral symmetry
in our formulation.
The last section is devoted to the summary and discussion.

\section{The conventional derivation from honeycomb lattice}\label{sec:conv}
\begin{figure}
\centering
\includegraphics[width=10cm,bb=0 2 585 452,clip]{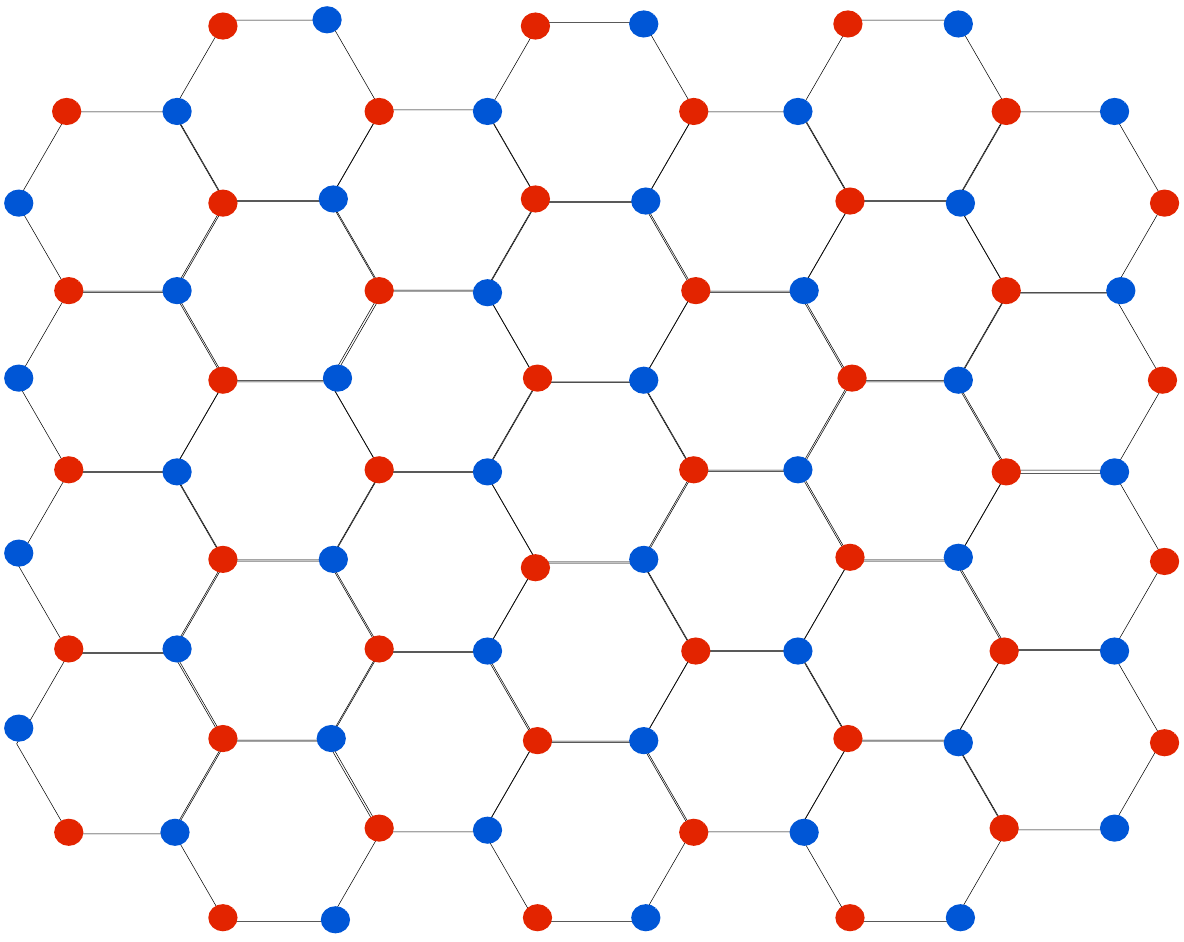}
\caption{
Honeycomb lattice is constituted of two triangular sub-lattices A and B, 
which are colored with red and blue respectively.
}\label{fig:honey}
\end{figure}

%
We first review the conventional derivation of 
Dirac fermion formulation from tight binding model of honeycomb lattice
\cite{Semenoff:1984dq}.
Let us start from the tight binding Hamiltonian
\begin{eqnarray}
\mathcal{H}
&=&
-t\sum_{\vec{r}\in A}\sum_{i=1,2,3}\sum_{\sigma}
\big[a^{\dagger}_{\sigma}(\vec{r})b_{\sigma}(\vec{r}+\vec{s}_{i})
    +b^{\dagger}_{\sigma}(\vec{r}+\vec{s}_{i})a_{\sigma}(\vec{r})\big]
\nonumber\\
&\ &
-t^{\prime}\Big[
\sum_{\vec{r}\in A}\sum_{j=1}^{6}\sum_{\sigma}a^{\dagger}_{\sigma}(\vec{r})a_{\sigma}(\vec{r}+\vec{s^{\prime}_{j}})
+\sum_{\vec{r}\in B}\sum_{j=1}^{6}\sum_{\sigma}b^{\dagger}_{\sigma}(\vec{r}+\vec{s^{\prime}_{j}})b_{\sigma}(\vec{r})
\Big],
\label{eq:hamiltonian}
\end{eqnarray}
where the first line is the nearest neighbor hopping term and 
the second line is the next-to-nearest neighbor hopping term,
and $t,t^{\prime}$ are hopping amplitudes.
$a(a^{\dagger})$ and $b(b^{\dagger})$ are the fermionic annihilation (creation)
operators of electrons on two triangular sublattices A and B respectively
(see in Figure \ref{fig:honey}).
$\vec{s}_{i}(i=1,2,3)$ and $\vec{s^{\prime}_{j}}(j=1,\cdots,6)$ 
denote the position vectors for three nearest neighbors
and the six next-to-nearest neighbors respectively.
$\vec{s}_{i}(i=1,2,3)$ is explicitly given in
\begin{eqnarray}
\vec{s}_{1}=a_{0}
\left(
\begin{array}{ccc}
1, &0
\end{array}
\right),\ \ 
\vec{s}_{2}=a_{0}
\left(
\begin{array}{ccc}
-1/2, &\sqrt{3}/2
\end{array}
\right),\ \ 
\vec{s}_{3}=a_{0}
\left(
\begin{array}{ccc}
-1/2, &-\sqrt{3}/2
\end{array}
\right),
\label{eq:basis}
\end{eqnarray}
where $a_{0}$ denotes a honeycomb lattice spacing.
We note that, in Graphene system, 
$t=2.8$ eV and $t^{\prime}=0.1$ eV \cite{Reich et al.},
and $a_{0}=1.42$ \AA\ \cite{Kotov:2010yh}.
In the following, we exclude spin index $\sigma$ for the sake of simplicity.
%
%
%
In order to find the Dirac points, 
we make Fourier transformation
\begin{eqnarray}
a(\vec{r})=\int\frac{d^{2}k}{(2\pi)^{2}}e^{i\vec{k}\cdot\vec{r}} \tilde{a}(\vec{k}), \ 
b(\vec{r})=\int\frac{d^{2}k}{(2\pi)^{2}}e^{i\vec{k}\cdot\vec{r}} \tilde{b}(\vec{k}),
\end{eqnarray}
for the fermionic creation and annihilation operator.
The nearest neighboring Hamiltonian represented 
in momentum space is given by
\begin{eqnarray}
\mathcal{H}=
\int\frac{d^{2}k}{(2\pi)^{2}}
\left(
\begin{array}{ccc}
\tilde{a}(\vec{k})\\
\tilde{b}(\vec{k})
\end{array}
\right)^{\dagger}
\left(
\begin{array}{ccc}
0&D(\vec{k})\\
D^{*}(\vec{k})&0
\end{array}
\right)
\left(
\begin{array}{ccc}
\tilde{a}(\vec{k})\\
\tilde{b}(\vec{k})
\end{array}
\right)
\label{eq:Ham_conv}
\end{eqnarray}
with
\begin{eqnarray}
D(\vec{k})=t\sum_{i=1,2,3}e^{i\vec{k}\cdot\vec{s}_{i}}.
\end{eqnarray}
Thus the energy eigenvalue of the above Hamiltonian is represented as
\begin{eqnarray}
E(\vec{k})=
\pm t\Big|\sum_{i=1,2,3}e^{i\vec{k}\cdot\vec{s_{i}}}\Big|. 
\end{eqnarray}
In the half-filled electron system, the negative and positive eigenvalues, 
which corresponds to the valence band and conduction band respectively,
appear, and there are two independent Dirac points
$\vec{K}_{\pm}$, in which $E(\vec{K}_{\pm})=0$ is fulfilled, 
on the fermi-surface.
\\
\\
In order to derive the low energy effective Hamiltonian,
we expand $D(\vec{k})$ around the Dirac points 
with respect to the momentum.
Regarding $\vec{K}_{\pm}$ and A, B site as spin degrees of freedom (DOF), and 
defining four component Dirac-spinor field $\tilde{\xi}(\vec{p})$ as
\begin{equation}
  \tilde{\xi}(\vec{p})=
  \big(\tilde{a}(\vec{K}_{+}+\vec{p}),\tilde{b}(\vec{K}_{+}+\vec{p}),
       \tilde{b}(\vec{K}_{-}+\vec{p}),\tilde{a}(\vec{K}_{-}+\vec{p})\big)^{T},
\end{equation}
the Hamiltonian in Eq.(\ref{eq:Ham_conv}) reads
\begin{eqnarray}
  \mathcal{H}\approx
  iv\sum_{i=1,2}\int_{\rm BZ}\frac{d^{2}p}{(2\pi)^{2}}
  \tilde{\xi}^{\dagger}(\vec{p})
  \big[\hat{\gamma}_{0}\hat{\gamma}_{i}p_{i}\big]\tilde{\xi}(\vec{p}).
\end{eqnarray}
Since the above is same form as the kinematic term of Dirac fermion field, 
$v= 3a_{0}t/2$ is interpreted as a fermi velocity of quasiparticles. 
Note that the gamma matrices $\hat{\gamma}_{0}, \hat{\gamma}_{1}, \hat{\gamma}_{2}$
satisfy Clifford algebra 
$\{\hat{\gamma}_{\mu},\hat{\gamma}_{\nu}\}=g_{\mu\nu}\cdot1_{4\times4}$,
where $g_{\mu\nu}$ is a metric in $2+1$ dimensional space-time.
Furthermore, introducing the matrix $\hat{\gamma}_{3}$, 
which is anti-commutative with
$\hat{\gamma}_{0}, \hat{\gamma}_{1}, \hat{\gamma}_{2}$,
we can define 
$\hat{\gamma}_{5}=i\hat{\gamma}_{0}\hat{\gamma}_{1}\hat{\gamma}_{2}\hat{\gamma}_{3}$ \cite{Gusynin:2007ix},
which we call as flavor-chiral symmetry
forbidding a (parity-invariant) mass term $m\tilde{\xi}^{\dagger}\hat{\gamma}_{0}\tilde{\xi}$.

We notice that, in the above derivation, it is not clear whether 
the theory is manifestly local, because each component of   fermion field is 
defined only in the subdomain near the low-energy points $K_{\pm}$
so that the continuity of the Dirac fermion in momentum space is not obvious.
In the next section, we will introduce 
an alternative derivation of Dirac fermion based on position space,
and also address 
the uniqueness of Dirac point and existence of physical modes. 

\section{Formulation in position space on honeycomb lattice}\label{sec:form}
%
\subsection{Tight binding model on the real space lattice}
%

\begin{figure}
\centering
\includegraphics[width=10cm,bb=0 2 585 452,clip]{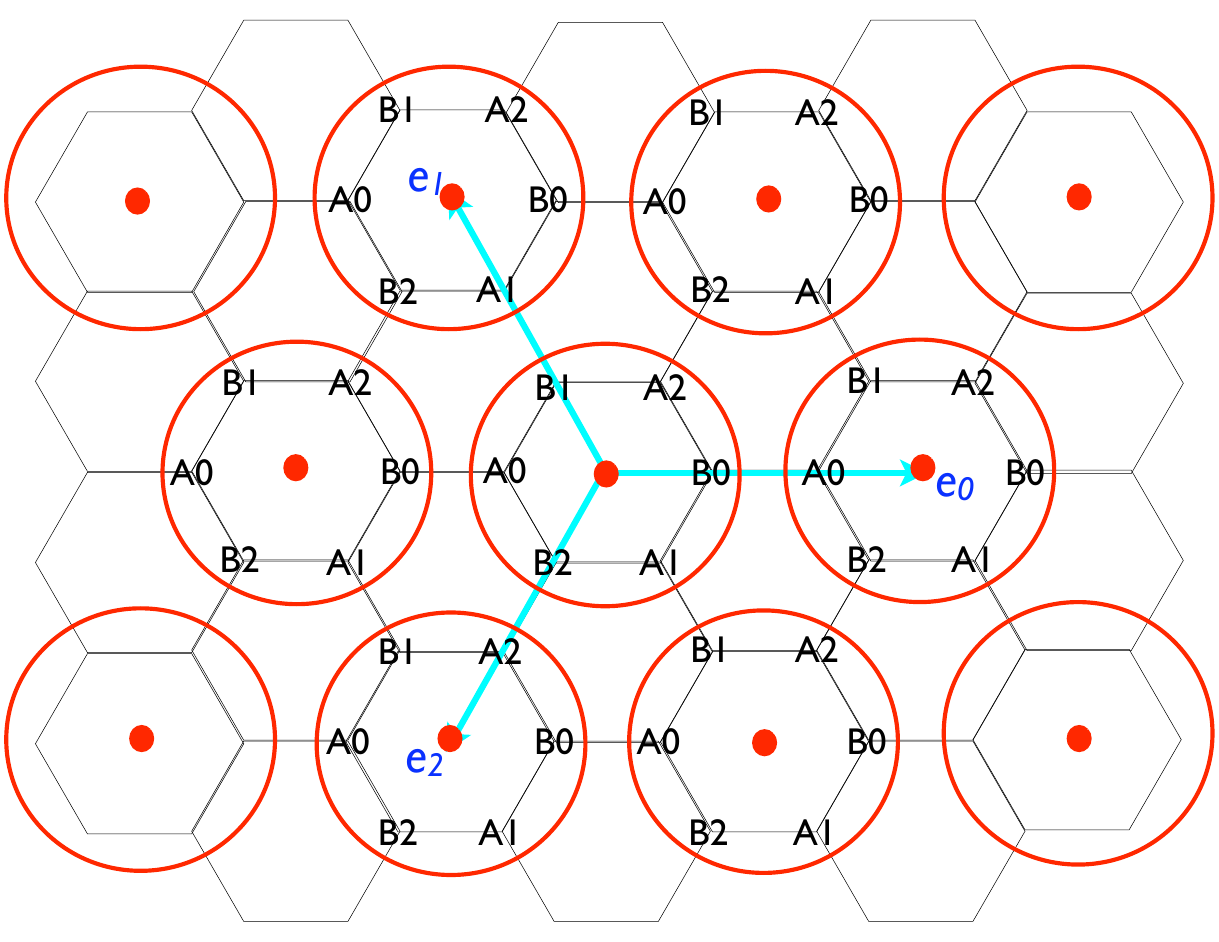}
\caption{
Honeycomb lattice}
\label{fig:honey2}
\end{figure} 

First we consider the new labeling of DOF of 
the fermionic creation and annihilation operator 
as shown in Figure \ref{fig:honey2}.
In this labeling, we define
$A\rho$ and $B\rho$ ($\rho=0,1,2$) as the new DOF having the 
operators on the site of honeycomb lattice. 
$\chi^{\dagger}_{I\rho}(\vec{x}),\chi_{I\rho}(\vec{x})$ are 
the new definition of creation and annihilation operators 
(the mass dimension of this operator is $\mathcal O(m)$.).
The arguments $\vec{x},\vec{y}$ are the positions
of  the center of hexagonal unit cell on the fundamental lattice,
\begin{eqnarray}
\vec{e}_{0}=a
\left(
\begin{array}{ccc}
1, &0
\end{array}
\right),\ \ 
\vec{e}_{1}=a
\left(
\begin{array}{ccc}
-1/2, &\sqrt{3}/2
\end{array}
\right),\ \ 
\vec{e}_{2}=a
\left(
\begin{array}{ccc}
-1/2, &-\sqrt{3}/2
\end{array}
\right),
\end{eqnarray}
where $a$ is the new lattice spacing 
defined as a distance between hexagonal unit cells.
The triangular sublattice $I(=A,B)$ of honeycomb lattice is 
composed of hexagonal unit cells bounded by red circles
in Figure \ref{fig:honey2}.
Note that the summation of three unit vectors vanishes as
$\vec{e}_{0}+\vec{e}_{1}+\vec{e}_{2}=0$
\footnote{
Here we note that there is following relation
between $\vec{e}_{\rho} (\rho=0,1,2)$ and $\vec{s}_{i} (i=1,2,3)$:
\begin{eqnarray}
\vec{e}_{0}=3\vec{s}_{1},\ 
\vec{e}_{1}=3\vec{s}_{2},\ 
\vec{e}_{2}=3\vec{s}_{3}.
\end{eqnarray}
}

Thus, in our formulation, the tight-binding Hamiltonian is expressed as
\begin{eqnarray}
  \mathcal{H}=
  \sum_{\vec{x},\vec{y}}
  \sum_{\rho,\rho^{\prime}}
  \left(
  \begin{array}{ccc}
    \chi_{A\rho}(\vec{x})\\
    \chi_{B\rho}(\vec{x})
  \end{array}
  \right)^{\dagger}
  \left(
  \begin{array}{ccc}
    t^{\prime}\Pi(\vec{x},\vec{y})_{\rho\rho^{\prime}}
   &t\Phi(\vec{x},\vec{y})_{\rho\rho^{\prime}}\\
    t\Phi(\vec{x},\vec{y})_{\rho\rho^{\prime}}^{\dagger}
   &t^{\prime}\Pi(\vec{y},\vec{x})_{\rho\rho^{\prime}}
  \end{array}
  \right)
  \left(
  \begin{array}{ccc}
  \chi_{A\rho^{\prime}}(\vec{y})\\
  \chi_{B\rho^{\prime}}(\vec{y})
  \end{array}
  \right),
  \label{eq:ham2}
\end{eqnarray}
where $\Phi(\vec{x},\vec{y})$ and $\Pi(\vec{x},\vec{y})$ are 
$3\times3$ matrix, 
\begin{eqnarray}
  \Phi(\vec{x},\vec{y})&=&
  \left(
  \begin{array}{ccc}
    T_{0}&   1&   1\\
    1&   T_{1}&   1\\
    1&   1&   T_{2}
  \end{array}
  \right)_{\vec{x},\vec{y}},\\
  \Pi(\vec{x},\vec{y})&=&
  \left(
  \begin{array}{ccc}
    0&   1+T_{0}+T_{1}^{\dagger}&   1+T_{0}+T_{2}^{\dagger}\\
    1+T_{0}^{\dagger}+T_{1}&   0&   1+T_{1}+T_{2}^{\dagger}\\
    1+T_{0}^{\dagger}+T_{2}&   1+T_{1}^{\dagger}+T_{2}&   0
  \end{array}
  \right)_{\vec{x},\vec{y}},
\end{eqnarray}
with backward shift in $\vec e_\rho$ direction and unit matrix,
\begin{eqnarray}
(T_{\rho})_{\vec{x},\vec{y}}=\delta_{\vec{x},\vec{y}+\vec{e}_{\rho}},\ 
(1)_{\vec{x},\vec{y}}=\delta_{\vec{x},\vec{y}}.
\end{eqnarray}
Now we define the forward shift
as $(T_\rho^\dag)_{\vec x,\vec y} = \delta_{\vec{x},\vec{y}-\vec{e}_{\rho}}$.

We note that the Hamiltonian in Eq.(\ref{eq:ham2}) is rewritten as
\begin{eqnarray}
\mathcal{H}
&=&
  a^2\sum_{\vec{x},\vec{y}}
    \chi(\vec{x})^{\dagger}
    \Big[tH(\vec{x},\vec{y})+t^{\prime}H^{2}(\vec{x},\vec{y})-3t'\Big]
    \chi(\vec{y}),
\label{eq:NNNH}
\end{eqnarray}
where $\chi(\vec{x})$ is labeled by two indices as $\chi_{I\rho} (I=A,B; \rho=0,1,2)$.
Using $\tau_{\pm}=(\tau_{1}\pm\tau_{2})/2$, which is 
\begin{eqnarray}
\tau_{+}=
\left(
\begin{array}{cc}
  0&   1\\
  0&   0
\end{array}
\right),
\tau_{-}=
\left(
\begin{array}{cc}
  0&   0\\
  1&   0
\end{array}
\right),
\end{eqnarray}
with Pauli matrices $\tau_{i}$ $(i=1,2,3)$,
$H$ is simplified as 
\begin{eqnarray}
H(\vec{x},\vec{y}) =
\tau_+ \otimes   \Phi(\vec{x},\vec{y})
 + \tau_- \otimes    \Phi^{\dagger}(\vec{x},\vec{y}).
\label{eq:hamil}
\end{eqnarray}
In the above equation, 
the former matrix in the tensor product acts on sub-lattice space $I=A,B$ 
while the latter acts on flavor space $\rho=0,1,2$.
$\Phi$ having index of flavor space is $3\times 3$ matrix, 
\begin{equation}
\Phi(\vec{x},\vec{y}) = (M - I_{3\times 3})\delta_{\vec x,\vec y}  
+ \sum_{\rho=0}^2\Gamma_\rho T_\rho(\vec x,\vec y),
\end{equation}
with 
\begin{eqnarray}
M=
\left(
\begin{array}{ccc}
  1&   1&   1\\
  1&   1&   1\\
  1&   1&   1
\end{array}
\right),\ 
\Gamma_{0}=
\left(
\begin{array}{ccc}
  1&   0&   0\\
  0&   0&   0\\
  0&   0&   0
\end{array}
\right),\ 
\Gamma_{1}=
\left(
\begin{array}{ccc}
  0&   0&   0\\
  0&   1&   0\\
  0&   0&   0
\end{array}
\right),\ 
\Gamma_{2}=
\left(
\begin{array}{ccc}
  0&   0&   0\\
  0&   0&   0\\
  0&   0&   1
\end{array}
\right).
\end{eqnarray}
Since the last term of Eq.(\ref{eq:NNNH}) merely shifts 
the origin of the energy, this term does not affect the dynamics at all.
Thus we neglect the constant term in the following discussion.
%
%
Defining the first and second derivative operators in $\vec{e}_\rho$ 
directions on the fundamental lattice as
\begin{eqnarray}
\nabla_\rho  = \frac{1}{2}(T_\rho^\dagger - T_\rho), \quad
\Delta_\rho  = \frac{1}{2}(T_\rho+ T_\rho^\dagger -2),
\end{eqnarray}
$\Phi$ is written as
\begin{eqnarray}
\Phi = M - \sum_\rho \Gamma_\rho \nabla_\rho 
+\frac{1}{2}\sum_{\rho}\Gamma_{\rho}\Delta_{\rho},
\end{eqnarray}
and, substituting the above equation into Eq.(\ref{eq:hamil}),
$H$ is also represented as
\begin{eqnarray}
H(\vec x,\vec y) =
\tau_{1}\otimes M\delta_{\vec x,\vec y}
-i\sum_{\rho}(\tau_{2}\otimes\Gamma_{\rho})\nabla_{\rho}(\vec x,\vec y)
+\frac{1}{2}\sum_{\rho}(\tau_{1}\otimes\Gamma_{\rho})\Delta_{\rho}(\vec x,\vec y),
\label{eq:realH}
\end{eqnarray}
Now the first and the second terms in Eq.(\ref{eq:realH}) 
are interpreted as the 
mass term and the kinetic term in the continuum limit ($a\rightarrow 0$), 
and also the third term vanishes, which is the second derivative term, 
in the continuum limit.
%
%

\subsection{Eigenvalue of the tight-binding Hamiltonian}
%
%
In this section, we discuss the eigenvalues of tight binding Hamiltonian 
$\mathcal{H}$ in Eq.(\ref{eq:NNNH}).
We consider the energy spectrum of the Hamiltonian Eq.(\ref{eq:NNNH}) 
in momentum space,
\begin{eqnarray}
\mathcal{H} = \int^{\pi/a}_{-\pi/a} \frac{d^2k}{(2\pi)^2} \tilde{\chi}^\dagger(\vec{k})
[t \tilde{H}(\vec{k})+t^\prime \tilde{H}^2(\vec{k})]\tilde{\chi}(\vec{k}),
\label{eq:modified ham}
\end{eqnarray}
where $\tilde{\chi}_{I\rho}(\vec{k})$ is Fourier representations of 
$\chi_{I\rho}(\vec{x})$, 
\begin{eqnarray}
\chi_{I\rho}(\vec{x}) =\int^{\pi/a}_{-\pi/a} 
  \frac{d^2k}{(2\pi)^2} e^{i\vec{k}\cdot\vec{x}} \tilde{\chi}_{I\rho}(\vec{k}),
\end{eqnarray}
and thus we have 
\begin{eqnarray}
 \tilde{H}(\vec{k}) &=& 
   \tau_1\otimes M + \sum_\rho(\tau_2\otimes \Gamma_\rho)\sin k_\rho
   + \sum_\rho (\tau_1\otimes\Gamma_\rho)(\cos k_\rho - 1),\\
 \tilde H^2(\vec k) &=& 
   1\otimes \Big[ 3M + \sum_\rho\{M,\Gamma_\rho\}(\cos k_\rho-1)
   - 2\sum_\rho\Gamma_\rho(\cos k_\rho-1)\Big]\nonumber\\
 &+& i\tau_{3}\otimes\sum_{\rho}[M,\Gamma_{\rho}]\sin k_{\rho},
\end{eqnarray}
with $k_\rho = \vec k\cdot \vec e_\rho$.

In order to give an intuitive picture for the eigenmodes of the Hamiltonian, 
we first consider the low energy limit, $k_\rho\rightarrow 0$,
where the $\chi$ approaches to the constant field.
In this limit, the Hamiltonian is  
\begin{eqnarray}
  H^{\rm low}\equiv
  \lim_{\vec k\rightarrow 0}\Big[t\tilde H(\vec k) + t^\prime \tilde H^2(\vec k)\Big] 
   = t(\tau_1\otimes M) + 3t^\prime(1\otimes M), 
\end{eqnarray}
and using 
\begin{eqnarray}
  \tilde\chi_{I\rho}(\vec{k})=\frac{1}{\sqrt{3}}\sum_{\rho^{\prime}=0,1,2}
            e^{i2\pi\rho\rho'/3}\psi_{I\rho^{\prime}}(\vec{k}),
  \label{eq:psi}
\end{eqnarray}
one easily sees the diagonalized form 
\begin{eqnarray}
  H^{\rm low}
  = t(\tau_1\otimes M^{\rm diag}) + 3t^\prime(1\otimes M^{\rm diag}),
  \label{eq:hlow}
\end{eqnarray}
with 
\begin{eqnarray}
M^{\rm diag}=
\left(
\begin{array}{ccc}
  3&   0&   0\\
  0&   0&   0\\
  0&   0&   0
\end{array}
\right).
\end{eqnarray}
This implies that the constant mode can be decomposed into 
two massless modes and one massive mode.

\begin{figure}
\centering
\vspace{-4.5cm}
\includegraphics[width=13cm,bb=0 2 585 452,clip]{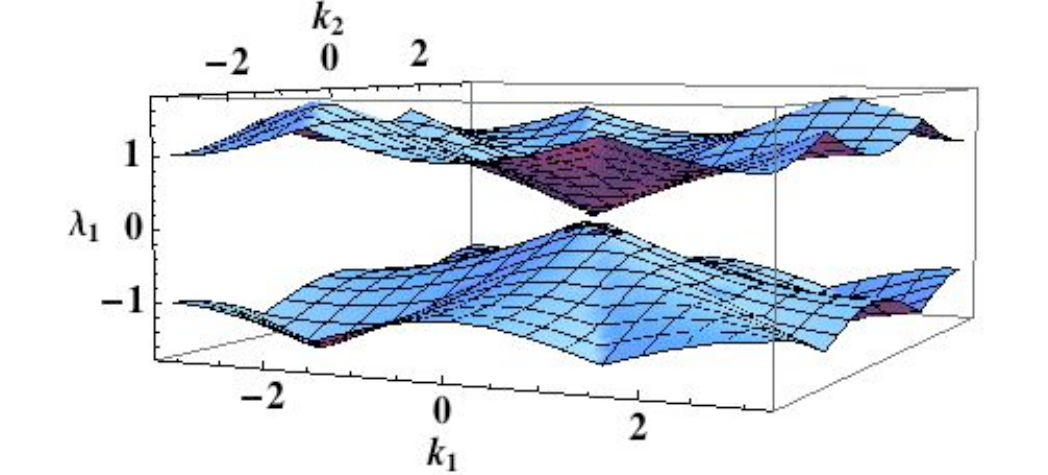}\\
\vspace{-4.5cm}
\includegraphics[width=13cm,bb=0 2 585 452,clip]{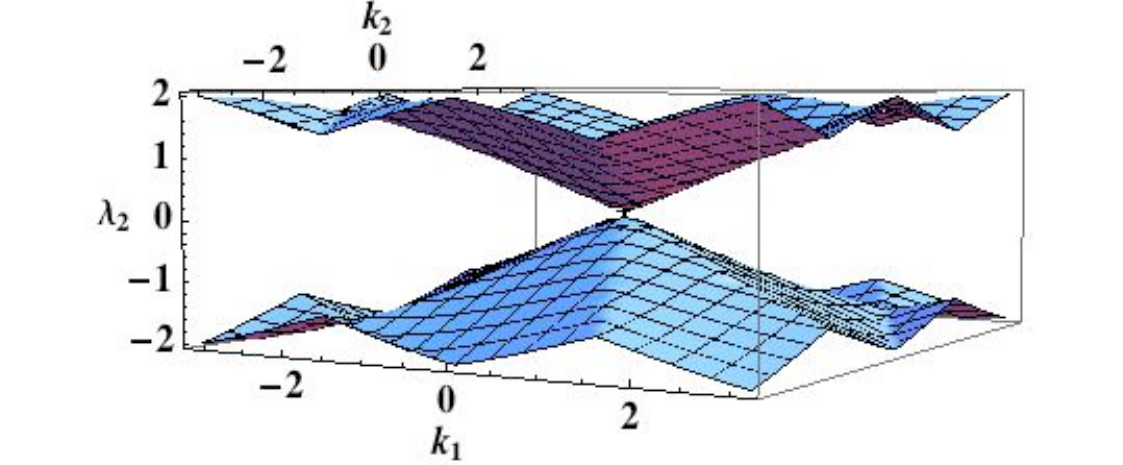}\\
\vspace{-4.5cm}
\includegraphics[width=13cm,bb=0 2 585 452,clip]{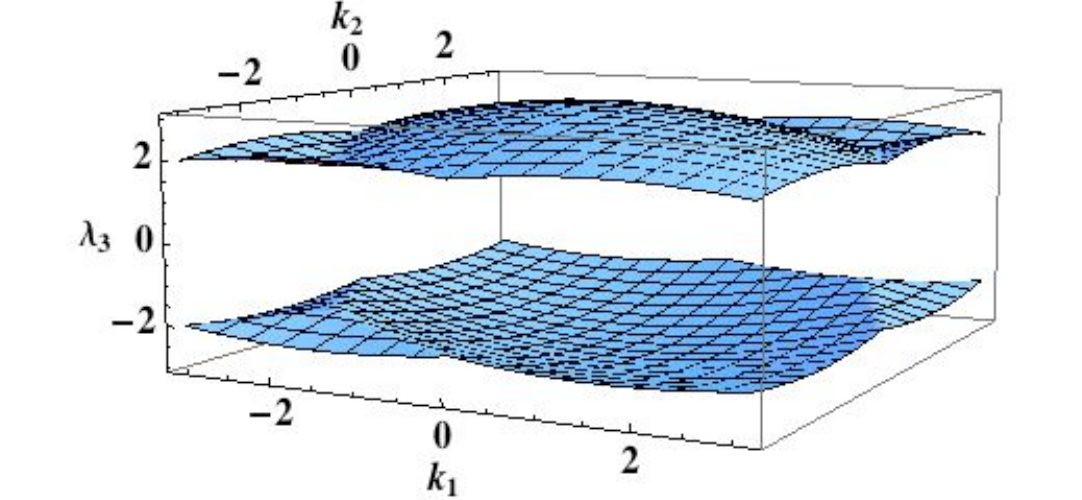}
\caption{Dispersion relation in the nearest-neighbor tight-binding model. Top, middle and bottom panels 
show $\lambda_{1}=\pm\phi_{1}(\vec{k})$, $\lambda_{2}=\pm\phi_{2}(\vec{k})$, and $\lambda_{3}=\pm\phi_{3}(\vec{k})$
respectively, where horizontal axes are $k_{1}=\vec{e}_{1}\cdot\vec{k}$,  $k_{2}=\vec{e}_{2}\cdot\vec{k}$.}
\label{fig:E123}
\end{figure}

Next, we investigate the Dirac point 
from the full energy spectrum of the Hamiltonian (\ref{eq:modified ham}).
Here we consider the eigenvalue equation when $t'=0$,
$\det(\lambda-\tilde{H})=0$, and,  
because the next-to-nearest neighboring term $\tilde H^2$ has
the same eigenvector with $\tilde H$, we easily
extends into $t'\ne 0$. 
After a simple algebraic calculation, we have 
\begin{eqnarray}
  \det(\lambda-\tilde{H}) 
  = \lambda^6 - 9 \lambda^4 -3(z_k+z_k^\ast-6)\lambda^2- |z_k-3|^2 =0
\label{eq:eigen}
\end{eqnarray}
with
\begin{eqnarray}
z_k= e^{-i\vec{k}\cdot\vec{e}_0}+e^{-i\vec{k}\cdot\vec{e}_1}+ e^{-i\vec{k}\cdot\vec{e}_2} .
\label{eq:zk}
\end{eqnarray}
Since Eq.(\ref{eq:eigen}) is a cubic equation for $\lambda^2$, 
the triple pair of energy eigenvalues of $\tilde{H}(\vec{k})$ 
should appear as $\pm \phi_1(\vec{k})$, $\pm \phi_2(\vec{k})$, $ \pm \phi_3(\vec{k})$, 
where $\phi_1,\phi_2,\phi_3$ ($0\leq \phi_i$) 
are functions of momentum $\vec{k}$ satisfied with Eq.(\ref{eq:eigen}).
One can easily see 
\begin{eqnarray}
\phi_1=\phi_2=0, & \phi_3=3,
\end{eqnarray}
at zero momentum as implied in Eq.(\ref{eq:hlow}). 
At finite momentum, the eigenvalues should be in the range of 
$0\le \phi_i\le 3$, and so that we define 
\begin{eqnarray}
  0 \leq \phi_1 \leq \phi_2 \leq \phi_3\leq 3.
\label{eq:range}
\end{eqnarray}
As a consequence, the eigenvalue of the Hamiltonian is expressed as
\begin{eqnarray}
  E^\prime_\pm(\phi_i) = \pm t \phi_i + t^\prime \phi_i^2 .
\label{eq:Ep}
\end{eqnarray}
Eq.(\ref{eq:eigen}) implies that the zero eigenvalue of $\tilde{H}(\vec{k})$ appears 
when $|z_k-3|^2=0$ holds. 
From Eq.(\ref{eq:zk}), it is obvious that this only takes place for $\vec{k}=0$. 
This means that the Dirac points uniquely appear at $\vec{k}=0$ 
in the BZ. 
This is a contrast to the traditional formulation, 
in which there are two Dirac points 
at the edge of the BZ (see section \ref{sec:conv}).
In position space formulation, Dirac fermion field 
possesses six DOFs, 
and, from the naive analysis of energy spectrum of 
the Hamiltonian, one sees that there are two physical modes and one massive mode. 
As a result of integrated out its massive mode, 
there remains four DOFs of physical mode, which is consistent 
with traditional one.
We will discuss more details later.  
In Fig.\ref{fig:E123}, we display the energy dispersion relation 
for eigenvalue of $H$ after exactly computed the solution 
of Eq.(\ref{eq:eigen}).
From this figure, we also figure out that there are two different 
dispersion relations associated with massive mode and physical mode. 

\begin{figure}
\centering
\includegraphics[width=8cm,bb=0 2 259 261,clip]{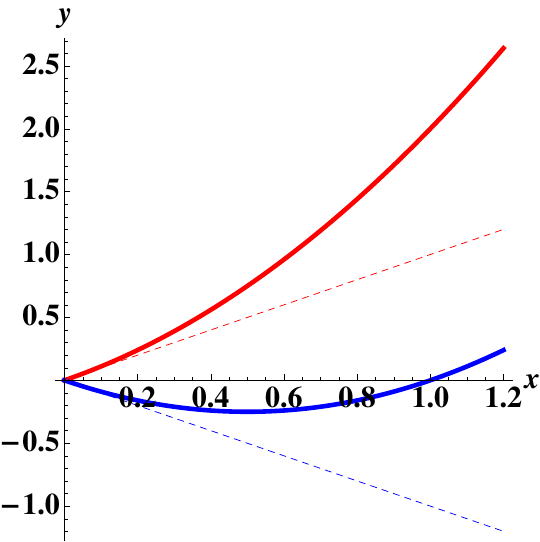}
\caption{
$x$ axis denotes $|\phi_{i}/\phi_{c}|$, and
$y$ axis denotes energy eigenvalues divided by $t^{\prime}\phi^{2}_{c}$.
The dashed lines colored with red and blue
are $E_{+}(\phi_{i})/t^{\prime}\phi^{2}_{c}=|\phi_{i}/\phi_{c}|$ and $E_{-}(\phi_{i})/t^{\prime}\phi^{2}_{c}=-|\phi_{i}/\phi_{c}|$ respectively.
While the solid lines colored with red and blue are
$E^{\prime}_{+}(\phi_{i})/t^{\prime}\phi^{2}_{c}=f_{+}(|\phi_{i}/\phi_{c}|)$ and $E^{\prime}_{-}(\phi_{i})/t^{\prime}\phi^{2}_{c}=f_{-}(|\phi_{i}/\phi_{c}|)$.
}\label{fig:eigenvalue}
\end{figure}

Here we discuss the effect of the next-to-nearest neighboring 
term into energy eigenmodes and dispersion relation. 
Eq.(\ref{eq:Ep}) is rewritten as 
\begin{eqnarray}
  E^\prime_\pm(\phi_i) = t^\prime \phi_c^2 f_\pm(|\phi_i/\phi_c|)
\label{eq:Ep2}
\end{eqnarray}
with $\phi_c=t/t^{\prime}$ and $f_\pm(x) = x^2 \pm x$.  
Taking $t^\prime=0$, the number of positive and negative energy eigenmodes 
is consistent, and 
it turns out that the fermi-surface for the half-filled electron system 
appears at zero energy level
(origin of dispersion relation in Figure \ref{fig:E123}).
However, taking account of the effect of the next-to-nearest neighbor hopping term 
$t^\prime\neq 0$, the situation is changed. 
Figure \ref{fig:eigenvalue}
shows that the negative eigenvalues $E^{\prime}_{-}(\phi_{i})$
remain in negative values unless $|\phi_{i}|$ exceeds $|\phi_{c}|$
which is crossing point of negative eigenvalue with zero, 
besides the eigenvalues $E^{\prime}_{+}(\phi_{i})$ stay in positive values 
at arbitrary $|\phi_{i}|$.
Thus, if $|\phi_{i}|$ does not exceed the threshold $|\phi_{c}|$,
the fermi surface remains in zero energy
due to the consistency between the number of 
the positive and negative eigenvalues.
On the other hand, if $|\phi_{i}|$ exceeds the threshold 
by choosing the abnormally large value of $t'$,
the fermi surface stays no longer in the same energy level.
In fact, because of $|\phi_{i}|\le 3$, 
the Dirac point stays at zero energy level even with 
the next-to-nearest neighbor hopping term  
as long as $3<|\phi_c|=|t/t^{\prime}|$.
In the Graphene, substituting the value of $t$ and $t'$
presented in \cite{Reich et al.}, 
since $|t/t^{\prime}|\simeq28$ is far from threshold, 
the fermi surface is not changed. 
%
%
%
%
%
%
%

\section{
The continuum limit
}\label{sec:continuum}
%
In this section, we consider the continuum limit and 
low-energy limit. 
Note that we ignore the higher order terms $\mathcal O(k^2)$
and $\mathcal O(a)$, and thus we set $t'=0$ in the following discussion.

In the momentum space, the tight binding Hamiltonian
in terms of the mass eigenstate,
as shown in Eq.(\ref{eq:hamil}) and (\ref{eq:modified ham}),  
is given as
\begin{eqnarray}
  \mathcal{H}&=&
  t \int^{\pi/a}_{-\pi/a}\frac{d^{2}k}{(2\pi)^{2}}
  \tilde{\psi}^\dagger(\vec{k})\big[
    \tau_{+}\otimes \tilde{\Phi}(\vec{k}) 
    + \tau_{-}\otimes \tilde{\Phi}^{\dagger}(\vec{k})\big]
\tilde \psi(\vec{k}),
\end{eqnarray}
where $\tilde{\psi}_{Ia}(\vec{k})$ was defined in Eq.(\ref{eq:psi}), 
and 
\begin{eqnarray}
\tilde{\Phi}(\vec{k})=
\frac{1}{3}
\left(
\begin{array}{ccc}
  b_{0}+b_{1}+b_{2}+6&   b_{0}+\omega^{2} b_{1}+\omega b_{2}& b_{0}+\omega b_{1}+\omega^{2} b_{2}\\
  b_{0}+\omega b_{1}+\omega^{2} b_{2}&   b_{0}+b_{1}+b_{2}-3& b_{0}+\omega^{2} b_{1}+\omega b_{2}\\
  b_{0}+\omega^{2} b_{1}+\omega b_{2}&   b_{0}+\omega b_{1}+\omega^{2} b_{2}&   b_{0}+b_{1}+b_{2}-3
  \end{array}
\right).
\end{eqnarray}
with $b_\rho=\exp(-i\vec{k}\cdot\vec{e}_\rho)$ ($\rho=0,1,2$).
Expanding $\tilde{\Phi}(\vec{k})$ with respect to $k$ up to 
$\mathcal O(\vec{k})$ as
\begin{eqnarray}
\tilde{\Phi}^{\prime}(\vec{k})&=&
\left(
\begin{array}{ccc}
  3&   0&   0\\
  0&   0&   0\\
  0&   0&   0
\end{array}
\right)
-\frac{a}{2}ik_{1}
\left(
\begin{array}{ccc}
  0&   1&   1\\
  1&   0&   1\\
  1&   1&   0
  \end{array}
\right)
-\frac{a}{2}ik_{2}
\left(
\begin{array}{ccc}
  0&   -i&   i\\
  i&   0&   -i\\
  -i&   i&   0
  \end{array}
\right)
+\mathcal{O}(k^2),
\end{eqnarray}
where $k_{i}(i=1,2)$ are components of momentum in Cartesian coordinates,
and integrating out the massive mode $\tilde{\psi}_{I0}(\vec{k})$ 
by its equation of motion
\begin{eqnarray}
\left(
\begin{array}{cc}
  0&   \tilde{\Phi}^{\prime}(\vec{k})_{00}\\
 \tilde{\Phi}^{\prime\dagger}(\vec{k})_{00}&   0
\end{array}
\right)
\left(
\begin{array}{cc}
\tilde{\psi}_{A0}(\vec{k})\\
\tilde{\psi}_{B0}(\vec{k})
\end{array}
\right)=
-\sum_{a=1,2}
\left(
\begin{array}{cc}
  0&   \tilde{\Phi}^{\prime}(\vec{k})_{0a}\\
 \tilde{\Phi}^{\prime\dagger}(\vec{k})_{0a}&   0
\end{array}
\right)
\left(
\begin{array}{cc}
\tilde{\psi}_{Aa}(\vec{k})\\
\tilde{\psi}_{Ba}(\vec{k})
\end{array}
\right),
\end{eqnarray}
the Hamiltonian is reduced to the following form
\begin{eqnarray}
&& \mathcal{H}_{\text{eff}} = \int^{\pi/a}_{-\pi/a}\frac{d^{2}k}{(2\pi)^{2}}
\tilde{\psi}^{\dagger}(\vec{k})
\Big[ v \Big\{
 k_{1}(\tau_{2}\otimes\sigma_{1})+k_{2}(\tau_{2}\otimes\sigma_{2})
\Big\} + \mathcal{O}(k^2)\Big]
\tilde{\psi}(\vec{k})
,\label{eq:ham_dirac}\\
&&\tilde{\psi}(\vec{k})=
\left(
\begin{array}{cccc}
\tilde{\psi}_{A1}(\vec{k})&
\tilde{\psi}_{A2}(\vec{k})&
\tilde{\psi}_{B1}(\vec{k})&
\tilde{\psi}_{B2}(\vec{k})
\end{array}
\right)^{T},
\end{eqnarray}
with $v=at/2$.
In the above tensor product representations, 
the latter of tensor structure acts on flavor space of physical mode $a=1,2$.

In the continuum limit,
the effective Hamiltonian has the following 4 global symmetries,
\begin{eqnarray}
1_{2\times2}\otimes1_{2\times2},\ 
\tau_{1}\otimes\sigma_{3},\ 
\tau_{2}\otimes1_{2\times2},\ 
\tau_{3}\otimes\sigma_{3}.
\end{eqnarray}
Now we consider the existence of parity invariant mass term in the Hamiltonian. 
This term is invariant under the parity transformation (which is exchange
symmetry of $A\leftrightarrow B$ and $x\rightarrow -x$ but $\rho\rightarrow\rho$
in Figure \ref{fig:honey2}), and so that we have 
\begin{eqnarray}
 m\tilde{\psi}^{\dagger}(\vec{k})
 (\tau_{1}\otimes1_{2\times2})\tilde{\psi}(\vec{k}).
 \label{eq:mass}
\end{eqnarray}
This is invariant under 
two global symmetries, $1_{2\times2}\otimes1_{2\times2},\ \tau_{1}\otimes\sigma_{3}$, 
whereas, under symmetry generated by 
$\tau_{2}\otimes1_{2\times2}$, $\tau_{3}\otimes\sigma_{3}$, 
this mass term is not invariant. 
Therefore, in analogy to QCD, we shall call the symmetry 
with generator $\tau_{2}\otimes1_{2\times2}$, $\tau_{3}\otimes\sigma_{3}$
as "flavor-chiral symmetry". 
Under this global symmetry, the parity invariant mass term Eq.(\ref{eq:mass})
is forbidden up to the first order of $\vec{k}$ (see in Table \ref{tab:symm}).
We notice that the higher derivative term than $\mathcal O(k^2)$
violates "flavor-chiral symmetry" similar to Wilson fermion.
It seems that the Parity invariant mass term may be induced 
through quantum corrections, which is associated with higher momentum effect, 
when interactions between electrons-electron and electron-photon
are turned on.

In the continuum limit, there exists a global flavor-chiral symmetry 
generated by $\tau_{3}\otimes\sigma_{3}$, however, 
as in the case of overlap fermion in lattice QCD, 
such global symmetry may be deformed by lattice artifact
at finite lattice spacing.
In the next section, we consider a possibility of flavor-chiral symmetry 
on position space formulation in honeycomb lattice.

Note that making Legendre transformation of Eq.(\ref{eq:ham_dirac}), we also derive 
the Lagarangian 
\begin{eqnarray}
\mathcal{L}=
i\bar{\psi}(t,\vec{x})\Big[
\partial_{0}\gamma_{0}-v\sum_{i=1,2}\gamma_{i}\partial_{i}
\Big]\psi(t,\vec{x})
\end{eqnarray}
where $\bar{\psi}=\psi^{\dagger}\gamma_{0}$ and 
gamma matrices $\gamma_{0},\ \gamma_{1},\ \gamma_{2}$ are,
\begin{eqnarray}
\gamma_{0}=
\left(
\begin{array}{cc}
  0&   1   \\
  1&   0
\end{array}
\right),\ 
\gamma_{1}=
\left(
\begin{array}{cc}
  -i\sigma_{1}&   0   \\
  0&   i\sigma_{1}
\end{array}
\right),\ 
\gamma_{2}=
\left(
\begin{array}{cc}
  -i\sigma_{2}&   0   \\
  0&   i\sigma_{2}
\end{array}
\right)
\end{eqnarray}
(in details see also appendix).
Apparently, these gamma matrices $\gamma_{0},\gamma_{1},\gamma_{2}$ 
satisfy Clifford algebra $\{\gamma_{\mu},\gamma_{\nu}\}=g_{\mu\nu}$,
where $g_{\mu\nu}$ is a metric in 2+1 dimensional space-time.
These gamma matrices are consistent with $\hat{\gamma}_{\mu}$ 
in section \ref{sec:conv}
by performing unitary transformation.

\begin{table}
\caption{Symmetry in effective theory for parity invariant mass term.}
\label{tab:symm}
\begin{center}
\begin{tabular}{|c|c|c|c|}\hline
\multicolumn{4}{|c|}{Global symmetry}\\ \hline
\multicolumn{2}{|c|}{preserved}  &  \multicolumn{2}{|c|}{broken}\\ \hline
\multicolumn{1}{|c|}{$1_{2\times2}\otimes1_{2\times2}$}  &  \multicolumn{1}{|c|}{$\tau_{1}\otimes\sigma_{3}$}  &  \multicolumn{1}{|c|}{$\tau_{2}\otimes1_{2\times2}$}  &  \multicolumn{1}{|c|}{$\tau_{3}\otimes\sigma_{3}$}\\
\hline
\end{tabular}
\end{center}
\label{default}
\end{table}

\section{Exact flavor-chiral symmetry}\label{sec:chiral}
%
In this section we employ the exact flavor-chiral symmetry
in position formulation in honeycomb lattice
to the next-to-leading order of tight binding approximation.

%
\begin{figure}[t]
\centering
\includegraphics[width=10cm,bb=0 2 755 585,clip]{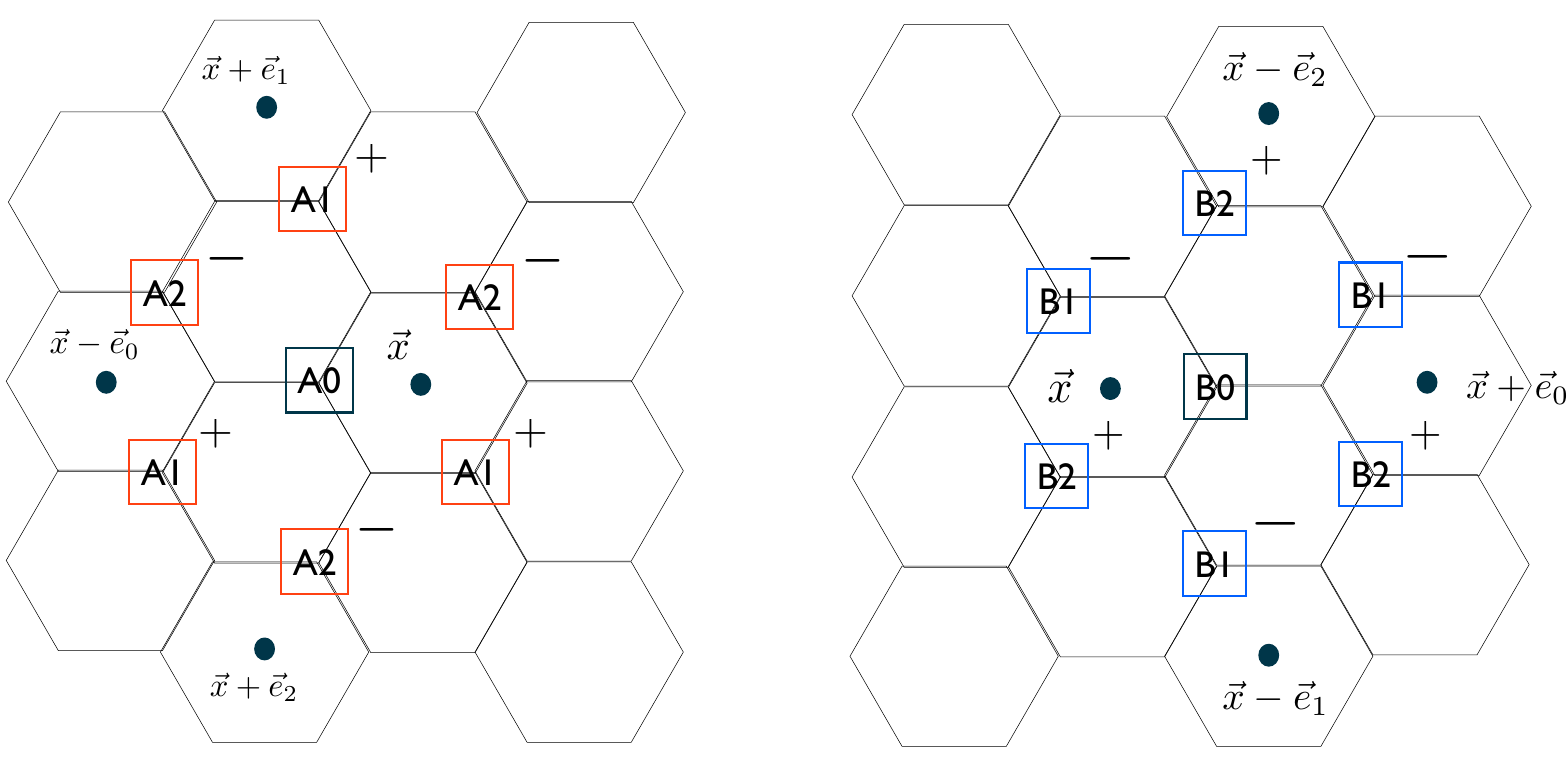}
\caption{
Geometrical picture of Eq.(\ref{eq:chiral transform A}), (\ref{eq:chiral transform B}).
Left and right panel show transformation for $\chi_{A0}(\vec{x})$ and one for $\chi_{B0}(\vec{x})$ respectively.
The transformation for $\chi_{A0}(\vec{x})$ ($\chi_{B0}(\vec{x})$) involves $\chi_{A\rho}(\vec{x})$ ($\chi_{B\rho}(\vec{x})$)
surrounded by red (blue) square, where sign denotes its overall factor.}
\label{fig:chiral}
\end{figure} 

First, we use the following ansatz for exact flavor-chiral symmetry of Hamiltonian
(\ref{eq:NNNH}), 
\begin{eqnarray}
\delta\chi(\vec{x})
&=&i\theta\Gamma_{5}\chi(\vec{x})
\nonumber\\
&=&i\theta\Big[
(\tau_{3}\otimes X)\chi(\vec{x})
+\frac{1}{2}\sum_{\rho}(\tau_{3}\otimes Y_{\rho})(\Delta_{\rho}\chi(\vec{x})+2\chi(\vec{x}))
+\frac{1}{i}\sum_{\rho}(1\otimes Z_{\rho})(\nabla_{\rho}\chi(\vec{x}))
\Big],
\nonumber\\
\label{eq:chiral transform}
\end{eqnarray}
where $X,Y_{\rho}$, and $Z_{\rho}$ are unknown $3\times3$ Hermitian matrices.
Based on this ansatz,
we determine the form of $X,Y_{\rho}$, and $Z_{\rho}$ from
the solution of symmetry equation $[\tilde H,\tilde\Gamma_{5}]=0$
in the momentum representation.
$\tilde\Gamma_5$ is defined as in momentum representation, 
which is consistent with generator $\tau_{2}\otimes 1_{2\times2}$
of global flavor-chiral symmetry in the continuum limit
\footnote{
There are two possibilities for $\tilde{\Gamma}_{5}$,
which are in agreement in the continuum limit.
One is $\tilde{\Gamma_{5}}=\tau_{3}\otimes X_{1}+\mathcal{O}(a)$,
and the other is $\tilde{\Gamma}_{5}=\tau_{2}\otimes X_{2}+\mathcal{O}(a)$.
However the latter is found not to satisfy the symmetry equation
at the second order of Taylar expansion around small momentum.
}.
We obtain the explicit forms of $X,Y_{\rho}$, and $Z_{\rho}$ as
\begin{eqnarray}
&&X=
\left(
\begin{array}{ccc}
  0&   -i&   i\\
  i&   0&   -i\\
  -i&   i&   0
\end{array}
\right),
\label{eq:coefficient1}
\\
&&Y_{0}=
\left(
\begin{array}{ccc}
  0&   -i&   i\\
  i&   0&   0\\
  -i&   0&   0
\end{array}
\right)
,\ 
Y_{1}=
\left(
\begin{array}{ccc}
  0&   -i&   0\\
  i&   0&   -i\\
  0&   i&   0
\end{array}
\right)
,\ 
Y_{2}=
\left(
\begin{array}{ccc}
  0&   0&   i\\
  0&   0&   -i\\
  -i&   i&   0
\end{array}
\right),
\label{eq:coefficient2}
\\
&&Z_{0}=
\left(
\begin{array}{ccc}
  0&   -1&   1\\
  -1&   0&   0\\
  1&   0&   0
\end{array}
\right)
,\ 
Z_{1}=
\left(
\begin{array}{ccc}
  0&   1&   0\\
  1&   0&   -1\\
  0&   -1&   0
\end{array}
\right)
,\ 
Z_{2}=
\left(
\begin{array}{ccc}
  0&   0&   -1\\
  0&   0&   1\\
  -1&   1&   0
\end{array}
\right).
\label{eq:coefficient3}
\end{eqnarray}
The details of the derivation are given in appendix  \ref{app:chiral}.

Next we consider how such flavor-chiral symmetry is interpreted as on honeycomb lattice.
Rewriting Eq.(\ref{eq:chiral transform}) in components of $\chi(\vec{x})$, 
the transformation for $\chi_{A\rho}(\vec{x})$,
$\chi_{B\rho}(\vec{x})$ reads
\begin{eqnarray}
\delta\chi_{A\rho}(\vec{x})&=&
\theta\big[
\chi_{A_{\rho+1}}(\vec{x}+\vec{e}_{\rho+1})
-\chi_{A_{\rho-1}}(\vec{x}-\vec{e}_{\rho})
\nonumber\\
& &
+\chi_{A_{\rho+1}}(\vec{x}-\vec{e}_{\rho})
-\chi_{A_{\rho-1}}(\vec{x}+\vec{e}_{\rho-1})
+\chi_{A_{\rho+1}}(\vec{x})
-\chi_{A_{\rho-1}}(\vec{x})],
\label{eq:chiral transform A}
\\
\delta\chi_{B\rho}(\vec{x})&=&
\theta\big[
-\chi_{B_{\rho+1}}(\vec{x}-\vec{e}_{\rho+1})
+\chi_{B_{\rho-1}}(\vec{x}+\vec{e}_{\rho})
\nonumber\\
& &
-\chi_{B_{\rho+1}}(\vec{x}+\vec{e}_{\rho})
+\chi_{B_{\rho-1}}(\vec{x}-\vec{e}_{\rho-1})
-\chi_{B_{\rho+1}}(\vec{x})
+\chi_{B_{\rho-1}}(\vec{x})].
\label{eq:chiral transform B}
\end{eqnarray}
%
%
%
One can see that the flavor-chiral transformation involves the next-to-nearest neighbor sites 
with alternating signs as in Figure \ref{fig:chiral}.
Using the conventional formulation
as in Eq.(\ref{eq:hamiltonian}),
the flavor-chiral transformation of $a(\vec{x})$, $b(\vec{x})$ is expressed as
\begin{eqnarray}
\delta a(\vec{x})=\theta[
a(\vec{x}+\vec{s}_{2}-\vec{s}_{3})-a(\vec{x}-\vec{s}_{1}+\vec{s}_{2})+a(\vec{x}+\vec{s}_{3}-\vec{s}_{1})
\nonumber\\
-a(\vec{x}-\vec{s}_{2}+\vec{s}_{3})+a(\vec{x}+\vec{s}_{1}-\vec{s}_{2})-a(\vec{x}-\vec{s}_{3}+\vec{s}_{1})]
\\
\delta b(\vec{x})=
\theta[
b(\vec{x}+\vec{s}_{2}-\vec{s}_{3})-b(\vec{x}-\vec{s}_{1}+\vec{s}_{2})+b(\vec{x}+\vec{s}_{3}-\vec{s}_{1})
\nonumber\\
-b(\vec{x}-\vec{s}_{2}+\vec{s}_{3})+b(\vec{x}+\vec{s}_{1}-\vec{s}_{2})-b(\vec{x}-\vec{s}_{3}+\vec{s}_{1})]
\end{eqnarray}
If we take a continuum limit $a\to0$, the above flavor-chiral transformation becomes 
$\delta\chi(\vec{x})=\theta[X+\sum_{\rho}Y_{\rho}]\chi(\vec{x})=3 i\theta X\chi(\vec{x})$.
In the mass basis, $X$ is transformed to the following form, 
\begin{eqnarray}
\left(
\begin{array}{ccc}
  0&   0&   0\\
  0&   1&   0\\
  0&   0&   -1
\end{array}
\right),
\end{eqnarray}
except for an overall factor.
Thus, the exact flavor-chiral symmetry corresponds to the global flavor-chiral symmetry 
$\tau_{3}\otimes\sigma_{3}$ as expected.
The flavor-chiral symmetry in the low energy effective theory has been discussed 
in some literature (see a review \cite{Gusynin:2007ix}).
In position space formulation, we also give 
the explicit formulation of flavor-chiral symmetry
at finite lattice spacing, which has, to our knowledge, 
not been known in previous literature.
It is known that axial $U(1)$ symmetry involves chiral anomaly in even dimensional space-time,
while in odd dimensional space-time chiral anomaly does not exist.
In this paper, we consider 2+1 dimensional fermion system, where time direction is continuous,
therefore there is no chiral anomaly and the flavor-chiral symmetry remains exact even at quantum level.
We comment that $\{\tau_{3}\otimes1_{3\times3},H\}=0$ has been often called 
as ``chiral symmetry'', for instance \cite{Ryu,Hatsugai},
in condensed matter physics, 
however, the exact flavor-chiral symmetry shown in this paper 
is characterized by symmetry equation to prevent the parity invariant mass term, 
and so that it is different from such definition.

\section{Summary and discussion}\label{sec:summary}
%
In this paper we present the construction of formulation of 
Dirac fermion from honeycomb lattice in position space.
In our formulation, we use the new labeling of fermion field
in which the fundamental lattice is composed of the centers of hexagonal unit cells.
The six sites in each unit hexagonal cell is reinterpreted as spin-flavor 
degrees of freedom.
Using this site-arrangement, 
the Hamiltonian in the nearest and the next-to-nearest 
neighboring term has kinetic term 
and second derivative term governing tensor structure 
with A , B site and three directions in hexagonal cell. 
In the analysis of energy spectrum, 
we show that one flavor has a mass of cutoff order 
and two quasiparticles are massless, and therefore, 
accounting for the degree of freedom of quasiparticle in position space formulation, 
it is consistent with momentum space formulation at all.
In our formulation, since the structure of the Dirac point is simplified, 
its uniqueness can be easily shown.
We also explicitly derive the global flavor-chiral symmetry at finite 
lattice spacing, which protects the masslessness of the Dirac fermion,
under the nearest neighboring interaction.

From the point of view in lattice gauge theory, 
the position space formulation corresponds to 
Staggered fermion formulation \cite{KlubergStern:1983dg}. 
We show that, starting from tight-binding model on honeycomb lattice, 
its Hamiltonian is represented as the tensor structure 
with the first derivative term and 
of Dirac fermion and the second derivative term, 
which correspond to kinematic term and lattice artifact respectively.
Regarding the degree of freedom of quasiparticle in honeycomb lattice 
as the flavor of Dirac fermion field in 2+1 dimension space-time, 
this formulation is in agreement with 
two-flavor massless staggered fermion formulation in hypercubic lattice.
In this case, the lattice spacing is defined as 
the distance between different unit hexagonal cell, and 
its physical point has been already known as finite value. 
This formulation provides a new picture as cut-off model 
for tight-binding approximation of the Graphene.

The position space formulation easily extends
toward the gauge interacting system. 
This also has the complementary information for understanding of
the connection with QED with 2+1 dimensional
fermion simulation \cite{Gorbar:2001qt,Shintani:2012sz}, 
Monte-Carlo simulation with electron-electron interaction 
\cite{Drut:2009,Armour:2010}
and honeycomb lattice simulation \cite{Brower:2012zd,Buividovich:2012nx}.
Furthermore, since our formulation has manifest structure of flavor symmetry,
it will be useful for implimentation of lattice simulations.

\section*{Acknowledgments}
%
The authors would like to thank Hidenori Fukaya, Yutaka Hosotani, and  Satoshi Yamaguchi for useful discussions.
 This work is supported by the Grant-in-Aid of the Japanese Ministry of Education (No. 20105002, 23105714(MEXT KAKENHI grant)).

\appendix
%

\section{Model in Lagrange formulation}
%
If we add a mass term to the effective theory, Hamiltonian is written as follows;
\begin{eqnarray}
\mathcal{H}=\int\frac{d^{2}k}{(2\pi)^{2}}
\tilde{\psi}^{\dagger}(\vec{k})[\alpha_{1}k_{1}+\alpha_{2}k_{2}+m\beta]\tilde{\psi}(\vec{k}),
\end{eqnarray}
where
\begin{eqnarray}
\alpha_{1}=
\left(
\begin{array}{ccc}
  0&   -i\sigma_{1}\\
  i\sigma_{1}&0
  \end{array}
\right),\ \ 
\alpha_{2}=
\left(
\begin{array}{ccc}
  0&   -i\sigma_{2}\\
  i\sigma_{2}&0
  \end{array}
\right).
\end{eqnarray}
$\beta$ is a Hermitian matrix and we may take following choices;
\begin{eqnarray}
\left(
\begin{array}{cc}
  0&   1   \\
  1&   0
\end{array}
\right),\ 
\left(
\begin{array}{cc}
  1&   0   \\
  0&   -1
\end{array}
\right).
\end{eqnarray}
Here the first gives parity even mass term and the second gives parity odd mass term.
However the parity odd mass term may be forbidden by parity, thus we choose the parity even mass term here.
Then transforming above Hamiltonian to Lagrangian in real space,
we obtain following Dirac Lagrangian in configuration space.
\begin{eqnarray}
\mathcal{L}&=&
i\psi^{\dagger}(t,\vec{x})\Big[
\partial_{0}+v\sum_{i=1,2}\alpha_{i}\partial_{i}-m\beta
\Big]\psi(t,\vec{x})\\
&=&
i\bar{\psi}(t,\vec{x})\Big[
\partial_{0}\gamma_{0}-v\sum_{i=1,2}\gamma_{i}\partial_{i}-m
\Big]\psi(t,\vec{x})
\end{eqnarray}
where $\bar{\psi}=\psi^{\dagger}\beta$ and $\gamma_{0}=\beta,\ \gamma_{1}=-\beta\alpha_{1},\ \gamma_{2}=-\beta\alpha_{2}$.
Evidently the gamma matrices $\gamma_{0},\gamma_{1},\gamma_{2}$ satisfy Clifford algebra $\{\gamma_{\mu},\gamma_{\nu}\}=g_{\mu\nu}$,
where $g_{\mu\nu}$ is a metric in 2+1 dimensional space-time.

\section{Explicit calculation of exact flavor-chiral symmetry}\label{app:chiral}
%
In order to determine $X,\ Y_{\rho},\ Z_{\rho}$, 
we employ momentum representation of $\chi(\vec{x}),\chi^{\dagger}(\vec{x})$
\begin{eqnarray}
\mathcal{H}
&=&
\int\frac{d^{2}k}{(2\pi)^{2}}
\tilde{\chi}^{\dagger}(\vec{k})\Big[
(\tau_{1}\otimes\Lambda)
+\sum_{\rho}e^{ik_{\rho}}(\tau_{-}\otimes\Gamma_{\rho})
+\sum_{\rho}e^{-ik_{\rho}}(\tau_{+}\otimes\Gamma_{\rho})
\Big]\tilde{\chi}(\vec{k})
\end{eqnarray}
with $\tau_{\pm}\equiv(\tau_{1}\pm i\tau_{2})/2$ and $\Lambda\equiv M-1$,
and for flavor-chiral transformation $\delta \tilde{\chi}(\vec{k})
=i\theta\tilde{\Gamma}_{5}(\vec{k})\tilde{\chi}(\vec{k})$
$\tilde{\Gamma}_{5}(\vec{k})$ is given as
\begin{eqnarray}
\tilde{\Gamma}_{5}(\vec{k})=
(\tau_{3}\otimes X)
+\sum_{\rho}e^{ik_{\rho}}\gamma_{\rho}
+\sum_{\rho}e^{-ik_{\rho}}\gamma_{\rho}^{\dagger},
\label{eq:gm5tilde}
\end{eqnarray}
with
\begin{eqnarray}
\gamma_{\rho}=\frac{\tau_{3}+1}{2}\otimes W^{\dagger}_{\rho}+\frac{\tau_{3}-1}{2}\otimes W_{\rho}.
\end{eqnarray}
$W_{\rho}$ is defined as $W_{\rho}=\frac{1}{2}(Y_{\rho}+iZ_{\rho})$.
Here, imposing $[\tilde{H}(\vec{k}),\tilde{\Gamma}_{5}(\vec{k})]=0$, we obtain following equations;
\begin{eqnarray}
&&\{\Lambda,X\}+\sum_{\rho}(\Gamma_{\rho}W_{\rho}+W_{\rho}^{\dagger}\Gamma_{\rho})=0\\
&&\{\Gamma_{\rho},X\}+\Lambda W_{\rho}^{\dagger}+W_{\rho}\Lambda=0\\
&&\Lambda W_{\rho}+W^{\dagger}_{\rho}\Lambda+\sum_{\sigma\neq\lambda(\sigma,\lambda\neq\rho)}(\Gamma_{\sigma}W_{\lambda}^{\dagger}+W_{\lambda}\Gamma_{\sigma})=0\\
&&\Gamma_{\rho}W_{\rho}^{\dagger}+W_{\rho}\Gamma_{\rho}=0\\
&&\Gamma_{\rho}W_{\sigma}+W_{\sigma}^{\dagger}\Gamma_{\rho}=0\ (\rho\neq\sigma).
\end{eqnarray}
Solving these equation for $X$ and $W_{\rho}(\rho=0,1,2)$, the solutions are found to be
\begin{eqnarray}
&&X=
\left(
\begin{array}{ccc}
  0&   -i&   i\\
  i&   0&   -i\\
  -i&   i&   0
\end{array}
\right),
\\
&&W_{0}=
\left(
\begin{array}{ccc}
  0&   -i&   i\\
  0&   0&   0\\
  0&   0&   0
\end{array}
\right)
,\ 
W_{1}=
\left(
\begin{array}{ccc}
  0&   0&   0\\
  i&   0&   -i\\
  0&   0&   0
\end{array}
\right)
,\ 
W_{2}=
\left(
\begin{array}{ccc}
  0&   0&   0\\
  0&   0&   0\\
  -i&   i&   0
\end{array}
\right),
\end{eqnarray}
where $Y_{\rho}$, $Z_{\rho}$ are given as $Y_{\rho}=W_{\rho}+W_{\rho}^{\dagger}$,
$Z_{\rho}=(W_{\rho}-W_{\rho}^{\dagger})/i$ respectively.

\input{ref}

\end{document}